\documentclass[aps,twocolumn,showpacs,showkeys,floatfix]{revtex4}

\usepackage{graphicx}
\usepackage{dcolumn}
\usepackage{bm}

\newcommand{\Tr}{\mbox{Tr}}
\newcommand{\grad}{\mbox{grad}}
\renewcommand{\div}{\mbox{div}}
\newcommand{\pdiffl}[2]{\frac{\partial #1}{\partial #2}}

\newcommand{\dfrac}[2]{\displaystyle\frac{#1}{#2}}

\begin{document}


\title{Shock and Release Temperatures in Molybdenum}

\date{March 12, 2007 -- LA-UR-07-1660}

\author{Damian C. Swift}
\email{dswift@lanl.gov}
\affiliation{%
   Group P-24, Physics Division, Los Alamos National Laboratory,
   MS~E526, Los Alamos, New Mexico 87545, USA
}

\author{Achim Seifter}
\affiliation{%
   Group P-24, Physics Division, Los Alamos National Laboratory,
   MS~E526, Los Alamos, New Mexico 87545, USA
}

\author{David B. Holtkamp}
\affiliation{%
   Group P-22, Physics Division, Los Alamos National Laboratory,
   MS~D410, Los Alamos, New Mexico 87545, USA
}

\author{David A. Clark}
\affiliation{%
   Group P-22, Physics Division, Los Alamos National Laboratory,
   MS~D410, Los Alamos, New Mexico 87545, USA
}

\begin{abstract}
Shock and release temperatures in Mo were calculated, taking account
of heating from plastic flow predicted using the Steinberg-Guinan model.
Plastic flow was calculated self-consistently with the shock jump conditions:
this is necessary for a rigorous estimate of the locus of shock states
accessible.
The temperatures obtained were significantly higher than predicted assuming
ideal hydrodynamic loading.
The temperatures were compared with surface emission spectrometry measurements
for Mo shocked to around 60\,GPa and then released into vacuum or into a LiF
window.
Shock loading was induced by the impact of a planar projectile, accelerated by
high explosive or in a gas gun.
Surface velocimetry showed an elastic wave at the start of release from the
shocked state;
the amplitude of the elastic wave matched the prediction to around 10\%,
indicating that the predicted flow stress in the shocked state was
reasonable.
The measured temperatures were consistent with the simulations,
indicating that the fraction of plastic work converted to heat was
in the range 70-100\%\ for these loading conditions.
\end{abstract}


\maketitle

\section{Introduction}
The behavior of matter subjected to extreme conditions through
dynamic loading is of interest from a direct physical standpoint,
as dynamic loading is often the only practical way to induce extreme
conditions, and because important engineering problems in hypervelocity
impact and weaponry involve dynamic loading
\cite{Eremets}.
Temperature is a key parameter in understanding the properties of matter,
as it governs the population of vibrational processes and excitations past
energy barriers.
Temperature is thus important for a physical understanding of many
types of behavior and the associated models.
The equation of state (EOS) includes contributions from the
excitation of atomic vibrations and electronic excitations.
Plastic flow is mediated by the excitation of dislocations and twin boundaries 
past Peierls barriers.
Phase changes depend on the thermodynamic state's location in the phase
diagram.
The kinetics of phase changes are described by 
the nucleation and growth of the daughter phase
in the matrix of the parent phase,
requiring the excitation of atoms past barriers.
Chemical reactions are governed by the excitation of atoms or electrons
over barriers.
Diffusion in condensed matter is the motion of atoms past the barriers
formed by their neighbors.
Conductivities include scattering contributions from thermal excitations.

Temperature is notoriously difficult to measure during dynamic loading,
in particular for opaque materials \cite{Luo07}.
Extreme states of matter are often hidden within a sheath of matter
in a different state -- this is the usual situation in shock loading
experiments.
Probes made of matter generally disrupt the state of interest,
e.g. by presenting an impedance mismatch to compression waves.
Much interesting physics in condensed matter occurs at compressions of
a few tens of percent, where the heating may be in the range of
a few hundred kelvin
and the resulting thermal emissions are small.
Most temperature measurements of shock-loaded systems 
have been made using photon emission spectroscopy,
commonly called pyrometry
\cite{Kormer65,Boslough89}. However, many
materials of interest (e.g. metals) are opaque in the relevant 
region of the spectrum: infra-red through visible for shocks up to
the terapascal regime.
Emission from an opaque material comes from the surface, which
cannot be maintained at the pressure of the initial shock for long
enough to allow useful emission spectra to be collected.
A transparent window can be placed in contact with the sample to maintain
an elevated pressure, but the mismatch in shock impedance must be
taken into account, along with the effect of heat conduction.
Accurate temperature data have been obtained from transparent sample materials,
where thermal radiation from the shocked state can escape from the
sample \cite{Luo_jgr_04,Luo_jpcm_04}.

Neutron resonance spectroscopy (NRS) has been investigated as a
fundamentally different technique for measuring the
temperature inside a dynamically-loaded specimen, which can be used on opaque
materials
\cite{Yuan05}.
Trial measurements of NRS temperatures were performed on shock-loaded Mo,
as a standard material for high pressure work;
the shock temperature was found to lie significantly above the
temperature predicted using reasonable EOS for Mo \cite{Yuan05}.
Measurements were also made using pyrometry of the temperature of Mo
which was shocked and then released into a LiF window or into vacuum;
these experiments also yielded temperatures which lay significantly
above EOS predictions \cite{Seifter04}.

Here we consider the effect of plastic flow on the Mo pyrometry
experiments.
Plastic flow was neglected in previous comparisons of predictions with
the temperature data.
The contribution of plastic work to heating has been mentioned in studies of
other metals \cite{Raikes79} 
but has not been quantified in detail or consistently.
The contribution to the total internal energy from plastic heating has
been estimated in order to extract the scalar EOS from shock data 
\cite{Morris80},
which involves a similar analysis of shock heating, though less general.

\section{Corrections and systematic uncertainties in pyrometry}
Polycrystalline materials, such as the Mo for which the discrepancy in
temperature measurements was reported, are heterogeneous in that they
are composed of an aggregate of grains of different crystallographic
orientation.
The Mo samples were machined from material which had been prepared by
pressing from powder, so there was a population of voids and there were
impurities concentrated along grain boundaries.
On shock loading and subsequent release, different regions of the sample
would thus respond differently, producing a variation in local temperature.
Given enough time, temperature variations disperse through thermal conduction,
but this typically takes of order microseconds for grains tens of micrometers
in size, which is long on the time scale of the experiments.
We wish to compare pyrometry measurements of temperature in Mo
with predictions using continuum models,
so the temperature of interest is the mean, bulk value.
Thermal radiation is described by the Stefan-Boltzmann relation,
where the total power is proportional to the fourth power of temperature.
Pyrometry measurements are often more accurate at shorter wavelengths
where the power varies with higher powers of temperature
\cite{Seifter_pyrowave_07}.
Thus unquantified temperature variations (spatial or temporal, within
the respective resolutions of the detectors) lead to an overestimate
of the mean temperature.
Spatial variation of brightness temperature has been observed in
shock-loaded Sn \cite{Seifter_07a}, which has a much lower flow stress than
Mo.

Pyrometry measurements from metals probe the surface temperature.
The surface is prone to increased plastic work around surface features such
as machining marks: when a rough surface in contact with material of
lower impedance (or vacuum, in the extreme case) is shocked,
flow may be exaggerated in pits and grooves.
At sufficiently high shock pressures, localized jetting may occur.
Any such localized increase in plastic flow will produce a higher local
temperature, which may appear as a higher mean temperature as discussed above.

For experiments in which the sample is observed through a transparent
window to maintain an elevated pressure,
there may be additional radiation from compression of any gas or glue in the
gap between the sample and the window, or from the shocked window material
itself.
These effects were considered and corrected for the Mo data
\cite{Seifter_07}.
After the shock passes from the sample into a window,
the temperature of the shock state is generally different to that
in the sample, so heat conduction can alter the temperature of
the sample surface.

In multiple channel pyrometry measurements, thermal emission from the sample
is recorded using a set of detectors responding to different ranges of
photon wavelength.
In the simplest case, a grey body spectrum can be fitted to the signals,
and the mean emissivity and temperature deduced from the shape of the spectrum.
In general, the emissivity of a material varies with state and wavelength,
and also with surface roughness, which may change during dynamic loading.
The emissivity may be measured directly, for example by ellipsometry.
It is more common to assume that
the multiple pyrometry channels over-sample the wavelength variation of
the radiance and emissivity, at least over part of the spectrum, 
allowing both to be deduced.
Again, these effects were taken into account for the Mo measurements
\cite{Seifter_07}.

Aside from gaps and glues between the sample and any window, 
it is common for components of the shocked target assembly to include 
sharp corners and low-density materials such as plastics, foams, and glues, 
as part of the engineering structure.
As with glued windows, low density materials in general may be shock-loaded
to a higher temperature than the sample.
Sharp internal corners, when shocked, may form jets with large amounts of plastic
heating.
Thermal emission from any of these sources may be present as a background
against which the emission from the sample must be distinguished.

The net effect of the complications associated with pyrometry measurements
on metals is that it is possible to over-estimate the temperature.

\section{Heating from plastic work in shock and release}
Shock compression involves the transit of a shock wave through each
element of material.
Subsequently, axial and lateral release and recompression waves reverberate
through the sample until it ultimately comes to rest at zero pressure.
In the simplest case, which many shock physics experiments are designed 
to achieve, the shock is steady and planar and the initial release is planar:
the material is compressed and released uniaxially.
In general, the shock and release may be converging or diverging, but locally the
compression and decompression induced is close to uniaxial.
Specifically, the strains are not not isotropic.
If a crystalline solid is subjected to non-isotropic strains
then shear stresses
must be induced, leading to plastic flow if the flow stress is exceeded.

In continuum dynamics situations such as shock loading,
simulations and analysis may be performed accurately by a scalar solution
of the shock jump and isentropic expansion relations if the material is
represented by an EOS, i.e. if the effects of elasticity and plastic flow
are ignored.
It is common practice to use spatially-resolving numerical simulations if
the material is to be represented with any greater complexity.
However, simulations of shock waves are complicated by the need to ensure that
the shock -- discontinuous at the continuum level -- is captured accurately
without inducing numerical artifacts such as oscillations or excess heating.
However, numerical methods have been developed to allow shock compression
and ramp decompression to be simulated by a scalar solution for general
material models including elasticity and plastic flow \cite{Swift_jcp_07}.
These numerical solutions were used to interpret the temperature measurements
on Mo.

Spatially-resolved simulations were also performed for comparison with
velocity history measurements.
These simulations used a Lagrangian representation of the shock experiments,
integrated in time with a predictor-corrector numerical scheme employing
artificial viscosity to stabilize the shock wave \cite{Benson92}.
The material models were identical with those used in the scalar solution.

The conservation equations for shock and release states in material dynamics
are usually formulated in terms of compression and pressure.
An order to take account of elastic-plastic effects, the
equations were formulated in terms of the stress and strain tensors.
Thus the Rankine-Hugoniot equations \cite{RH} for conservation across the shock
were expressed in terms of the total stress normal to the shock, $\tau_n$,
rather than the pressure $p$:
\begin{eqnarray}
u_s^2 & = & -v_0^2\dfrac{\tau_n-\tau_{n0}}{v_0-v}, \\
\Delta u_p & = & \sqrt{-(\tau_n-\tau_{n0})(v_0-v)}, \\
e & = & e_0 - \frac 12 (\tau_n+\tau_{n0})(v_0-v),
\end{eqnarray}
where $v$ is specific volume (the reciprocal of the mass density $\rho$),
$e$ is specific internal energy,
$u_s$ is the speed of the shock wave with respect to the
material, $\Delta u_p$ is the change in material
speed normal to the shock wave (i.e., parallel to its direction
of propagation), and subscript $0$ refers to the initial state.
The specific internal energy was defined to {\it exclude} elastic
strain energy, so the energy equation above included only the
volumetric and plastic strain contributions to the volume change.
The relation for adiabatic compression and release was expressed similarly:
\begin{equation}
\dot e = \left\{\begin{array}{ll}
   -\dfrac{p\div\,\vec u}\rho\quad&:\quad\mbox{elastic} \\
   \frac 1\rho\left(||\sigma\grad\,\vec u||-p\div\,\vec u\right)
   \quad&:\quad\mbox{plastic}
\end{array}\right.
\end{equation}
where $\tau$ is the stress tensor,
$\sigma$ the deviatoric stress
\begin{equation}
\sigma\equiv\tau-\frac 13\Tr\,\tau I=\tau+pI,
\end{equation} 
$\grad\,\vec u$ the velocity gradient
tensor, and $\div\,\vec u$ its trace.
For uniaxial compression,
$||\sigma\grad\,\vec u||=\sigma_n\partial \vec u_n/\partial r_n$
i.e. the product of the components in the direction normal to the wave,
all others being zero.
In the non-spatially-resolved calculations,
the velocity gradient was simply the assumed or imposed strain rate.

The state of the material was expressed in terms of $\rho$ and $e$
(allowing a mean pressure $p(\rho,e)$ to be calculated from the EOS),
a deviatoric elastic strain tensor $\epsilon$
(allowing the deviatoric stress contributions $\sigma$ to be calculated),
and a scalar equivalent plastic strain $\tilde\epsilon_p$,
used to calculate work hardening.
As discussed elsewhere \cite{Swift_jcp_07},
a hyperelastic formulation using strain rather than a hypoelastic formulation
using stress was preferred for consistency and accuracy in situations where
shear strains are applied at different compressions.
Thus the stress deviator $\sigma$ was calculated from the instantaneous strain,
\begin{equation}
\sigma = 2 G \epsilon,
\end{equation}
where $G(\rho,T)$ is the shear modulus.
Plastic flow was taken to occur using a von Mises yield surface \cite{Hill}.
Deformation was plastic rather than elastic if the scalar effective shear stress
\begin{equation}
\tilde\sigma\equiv\sqrt{f_\sigma ||\sigma^2||}
\end{equation}
exceeded the yield stress $Y(\rho,e,\tilde\epsilon_p)$,
in which case plastic strain for work hardening was accumulated at a rate
\begin{equation}
\dot{\tilde\epsilon}_p=
   \frac{f_\epsilon}2\frac{||\dot\epsilon\epsilon||+||\epsilon\dot\epsilon||}{\tilde\epsilon},
\end{equation}
where $\dot\epsilon$ is the deviatoric part of the symmetric part of the velocity
gradient,
\begin{equation}
\dot\epsilon\equiv \dot E-\frac 13\Tr\,\dot EI\quad:\quad \dot E\equiv\frac 12(U+U^T),\quad
U\equiv\grad\,\vec u.
\end{equation}
If $\tilde\sigma<Y$, the elastic deformation was simply $\dot\epsilon$.
For uniaxial compression along the $x$-direction, the only non-zero component
of $U$ is $[U]_{100}$.

If plastic flow occurs, then the material is always heated to some degree.
Plastic flow occurs through the motion and generation of defects in the
crystal lattice, such as dislocations.
Usually in polycrystalline materials, defects accumulate during plastic
deformation.
Heating generally represents less than the total plastic work
as some potential energy is absorbed in the structure of defects.
The fraction of plastic work converted to heat $f_p$ is thought to be 0.85-0.95.
It was assumed here to be 0.9.
Thus the contribution of plastic work to heating was
\begin{equation}
\dot e_p = f_p\frac{||\sigma \dot e||}\rho
\end{equation}
if $\tilde\sigma>Y$ and zero otherwise.

When a metal is deformed, shear strains result in the accumulation of
elastic energy until the flow stress is reached.
Continued deformation results in plastic work.
If the material work-hardens, the rate of plastic working increases.
If thermal softening occurs, the rate of plastic working decreases but
the stored elastic energy is also converted to plastic work.
In an idealized material exhibiting a constant flow stress 
(elastic-perfectly plastic), arbitrarily large amounts of plastic work may be
accumulated by large deformations -- uniaxial as well as pure shear -- beyond
the flow stress.
Ceramics may behave differently, the flow stress decreasing to a small 
fraction of its initial value as deformation continues beyond the elastic limit
\cite{Raikes79}, presumably as interatomic bonds are broken and brittle
damage occurs.
Many transparent materials are ceramic; this reduction of the flow stress
may explain why good agreement has been obtained between shock temperatures
and predictions neglecting heating from the constitutive response \cite{Luo_jpcm_04}.

Plastic flow is largely irreversible.
If a sample of material is shock loaded and then released \footnote{%
In purely hydrodynamic analyses, where the effect of plastic flow is
ignored, release from a shocked state follows an isentrope.
This is no longer true when additional dissipative processes occur,
such as plastic flow and viscosity, which lead to heating
with an increase in entropy.
The term `quasi-isentropic' is sometimes used in this context,
particularly for shockless compression;
here we prefer to refer to the release loci as adiabats since this is
a more specific term. 
},
the pressure reduces on release but further plastic work is done.

Mo was represented by an empirical EOS fitted to shock compression data
\cite{Steinberg96,SESAME,Greeff}, with a deviatoric strength model developed
and calibrated to data on the amplitude and shape of elastic waves running
ahead of shocks \cite{Steinberg96}.
The EOS was likely to be accurate to a few percent in temperature
for the shock pressures of a few tens of gigapascals considered here.
The Steinberg-Guinan strength model includes a prediction of the flow stress
at elevated pressures.
The flow stress, and hence the heating from plastic flow, was uncertain at the level of
a few tens of percent.
As discussed below, measurements of surface velocity provided an independent
measure of the flow stress.

Material models for continuum dynamics are often implemented in varying
ways in different computer programs.
The results may depend on details such as the way in which numerical limits,
e.g. on flow stress, are applied.
In our simulations, the EOS was represented by an expression for
pressure $p$ in terms of mass density $\rho$ and specific internal energy $e$.
This is sufficient to allow the dynamical equations for the continuum to be
integrated in time.
Two different EOS were used, a tabular form from the `SESAME' library
\cite{SESAME}, and an analytic form of the Gr\"uneisen type, using the
principal Hugoniot as the reference curve \cite{Steinberg96}:
shock speed $u_s$ in terms of particle speed $u_p$,
\begin{equation}
u_s=c_0 + s_1 u_p,
\end{equation}
together with a relation for the Gr\"uneisen parameter
\begin{equation}
\gamma(\rho)=\gamma_0 + b(\rho/\rho_0-1).
\end{equation}
The shear modulus $G$ and flow stress $Y$ followed the Steinberg-Guinan
model \cite{Steinberg80}, which includes explicit dependence on temperature $T$ and accumulated
plastic strain $\epsilon_p$:
\begin{eqnarray}
G(p,T) & = & G_0 \left[1+Ap(\rho/\rho_0)^{-1/3}-B(T-T_0)\right] \\
Y(p,T) & = & Y_0 f(\epsilon_p) G(p,T)/G_0 \\
f(\epsilon_p) & = & \mbox{min}\left[(1+\beta\tilde\epsilon_p)^n,Y_{\mbox{max}}/Y_0\right].
\end{eqnarray}
Because of the scaling of flow stress by shear modulus, the maximum flow stress
at high pressures was not limited by the `maximum' flow stress $Y_{\mbox{max}}$
-- this allowed the flow stress to be significantly greater than $Y_{\mbox{max}}$
in the Mo impact experiments.
The usual definition of the Steinberg-Guinan model \cite{Steinberg96}
includes an explicit initial plastic strain from manufacture;
we treated $\epsilon_p$ as a local material parameter in addition to
$\rho$, $e$, and the elastic strain, and set $\tilde\epsilon_p$ to a non-zero value
in the initial conditions if required.
The factors $f_\epsilon$ and $f_\sigma$ used in calculating the scalar effective magnitudes
of the corresponding tensors were chosen for consistency with the definitions of
stress and strain used in deducing strength parameters for Mo from experiments:
$f_\epsilon=f_\sigma=3/2$.

The SESAME EOS were defined as a pair of tables $\{p,e\}(\rho,T)$, so the $p(\rho,e)$
relation was obtained by numerical inversion and the temperature was readily
calculated.
Temperatures were calculated from the Gr\"uneisen EOS with reference to a compression 
curve along
which the temperature and specific internal energy were known, $\{T_r,e_r\}(\rho)$, 
and using
a specific heat capacity defined as a function of density $c_v(\rho)$
(constant in practice).
The reference curve chosen was the zero kelvin isotherm (`cold curve'), $T_r=0$\,K.
This was calculated from the principal isentrope $\left.e(\rho)\right|_{s_0}$ using
the estimated density variation of Gr\"uneisen parameter:
\begin{equation}
e_r(\rho) = \left.e(\rho)\right|_{s_0} - T_0 c_p e^{a(1-\rho_0/\rho)}
   \left(\frac\rho{\rho_0}\right)^{\gamma_0-a}.
\end{equation}
The isentrope was calculated by numerical integration of the second law
of thermodynamics,
\begin{equation}
\left.\pdiffl ev\right|_s=-p(1/v,e).
\end{equation}
Mechanical properties and temperatures calculated by either EOS gave the
same result to $o(1\%)$, which constitutes good agreement for models
in material dynamics.
The Gr\"uneisen EOS have slightly smoother loci, so the results presented
below are from this EOS.

Simulations were performed in units of millimeters, gigapascals,
microseconds, kelvin, and Mg/m$^3$=g/cm$^3$.
Parameters for Mo in these units are listed in Table~\ref{tab:Moparams}.

\begin{table}
\caption{Gr\"uneisen equation of state and Steinberg-Guinan strength
   parameters for Mo.}
\label{tab:Moparams}
\begin{center}
\begin{tabular}{|c|c|l||c|c|l|}\hline
\multicolumn{3}{|c|}{equation of state} &
\multicolumn{3}{|c|}{strength} \\ \hline
$\rho_0$ & 10.2 & g/cm$^3$ & $G_0$ & 125 & GPa \\
$c_0$ & 5.143 & km/s & $Y_0$ & 1.6 & GPa \\
$s_1$ & 1.255 & & $A$ & $1.14\times 10^{-2}$ & GPa$^{-1}$ \\
$\gamma_0$ & 1.59 & & $B$ & $1.52\times 10^{-4}$ & K$^{-1}$ \\
$b$ & 0.30 & & $\beta$ & 20 & \\
$c_p$ & $2.43\times 10^{-4}$ & MJ/kg.K & $n$ & 0.15 & \\
$a$ & 1.3 & & $Y_{\mbox{max}}$ & 2.8 & GPa \\
\hline\end{tabular}

Source: \cite{Steinberg96} with unit conversions.
\end{center}
\end{table}

\section{Pyrometry experiments on molybdenum}
\begin{figure}
\begin{center}\includegraphics[scale=0.35]{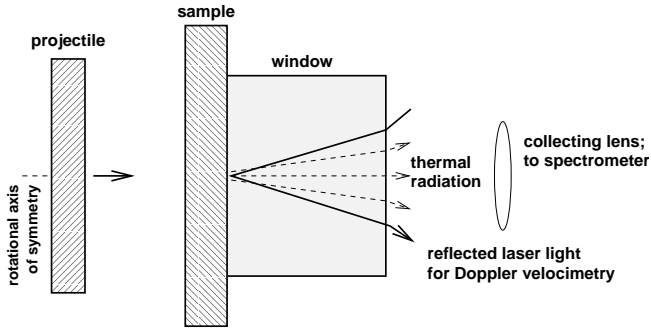}\end{center}
\caption{Schematic of impact-induced shock experiments with
   surface temperature measurements.
   Aspect ratios are representative of the experiments discussed here.
   If the window is omitted, the experiment measures the free surface
   (zero normal stress) temperature.}
\label{fig:impactschem}
\end{figure}

\begin{figure}
\begin{center}\includegraphics[scale=0.7]{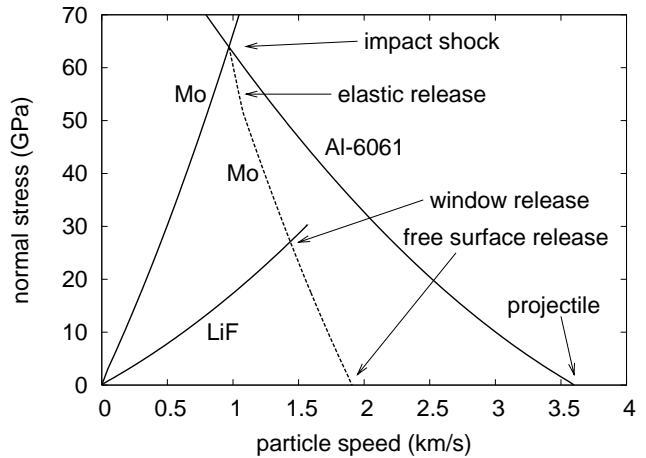}\end{center}
\caption{Shock and release states induced in impact experiments
   with and without a window.
   Solid lines: shock Hugoniots; dashed line: release adiabat.
   Example calculation for Al-6061 projectile traveling at 3.6\,km/s,
   impacting stationary Mo target, releasing into LiF window
   or into vacuum.
   The initial states of the Mo and LiF are at the origin;
   the initial state of the projectile is at zero normal stress and 3.6\,km/s.
   On impact, the shock states in the projectile and sample are
   at the elevated pressure intersection marked `impact shock.'
   When the shocked state in the Mo releases into the LiF window,
   the resulting state is the intersection marked `window release.'
   When the shocked state in the Mo is released at a free surface,
   the resulting state is the zero normal stress state marked
   `free surface release.'
   Release from the shocked state shows an inflexion when plastic flow occurs.}
\label{fig:uppcmp}
\end{figure}

Pyrometry measurements of the temperature in shocked and released Mo have been 
made using two types of experiment.
In both cases, the shock was induced by the impact of a flat projectile.
The projectiles were accelerated using a high explosive launcher,
as in the NRS experiments, and by a gas gun.
The pyrometry measurement was performed at the surface opposite the impact,
the shocked state releasing either to vacuum or into a LiF window
to sustain an elevated pressure
(Figs~\ref{fig:impactschem} and \ref{fig:uppcmp}).
In each case, the shock state was calculated using the published EOS
and strength properties for the projectile and Mo target.

In all cases, the impact conditions were calculated using the scalar
solution, and were repeated with and without strength in all components
of the impact experiment.
For experiments with a LiF window,
the temperature in the window was also predicted;
a high temperature would signal an increased possibility of thermal
radiation from the window obscuring the emissions from the Mo sample.
Where an uncertainty in impact velocity was reported, the calculations
were repeated for velocities at the extremes of uncertainty,
giving an estimate of the uncertainty in pressure and temperature.

Similar calculations were performed with and without strength
in each component separately.
The Steinberg-Guinan model is least appropriate for LiF, so this is
the only component where it would be useful to make such additional
comparisons.
However, the effect on states releasing into LiF were dominated by the 
strength of the Mo, so the additional comparisons are omitted for clarity.
An indication that the contribution of strength in the LiF is
a small effect in the simulations is that the predicted shock temperature
in LiF varied much less as a function of strength than did the temperature of 
any of the Mo states.

Taking strength into account, on release into LiF,
the normal stress in the Mo was lower than the in-plane stress
because the elastic strain is a distension in the axial direction.
For this reason, the calculations with strength have a lower normal stress:
a result which may be counterintuitive.

Various improvements could be made in future pyrometry measurements
to reduce the temperature uncertainties.
Some optimization could be performed by repeating experiments
multiple times,
adjusting detector gains and digitization ranges for best
accuracy.
However, the difficulty and cost of each projectile impact experiment
can make multiple repeats impractical.

\subsection{Gas gun}
The projectile was Ta, 3\,mm thick, accelerated to 1.70\,km/s 
using a two stage gas gun.
The target was Mo, 5\,mm thick.
Thermal emission was measured on release into a LiF window,
using a 7 channel pyrometer.
The measured release temperature was $683\pm 41$\,K.

\begin{figure}
\begin{center}\includegraphics[scale=0.7]{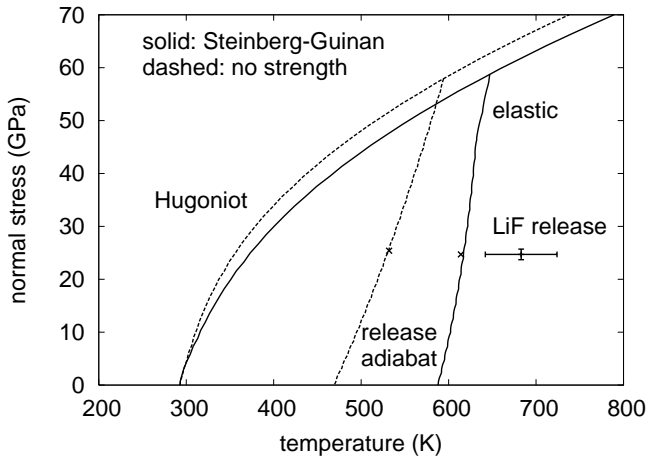}\end{center}
\caption{Temperature measurement from a shock of 59\,GPa,
   on release to 25\,GPa into LiF,
   compared with predictions based on the Steinberg-Guinan strength model
   and with strength neglected.
   The crosses on the release adiabats show where release pauses when
   a shock is transmitted into the LiF.
   When strength is included, the first portion of release is elastic;
   the elastic portion of the adiabat is marked;
   flow becomes plastic below the inflexion.}
\label{fig:tpcmpgun}
\end{figure}

The shocked state in the Mo was calculated to be 58.7\,GPa and 645\,K,
of which 51\,K was from plastic work.
The state on release into LiF was thus calculated to be
24.8\,GPa and 614\,K, of which 82\,K was from plastic work.
The measured surface temperature was just 1.5 standard deviations above the 
temperature predicted
using the Steinberg-Guinan strength model,
and more than three standard deviations above the temperature predicted
ignoring material strength.
(Table~\ref{tab:states} and Fig.~\ref{fig:tpcmpgun}.)

\subsection{Forest Flyer}
The high explosive launcher used the Forest Flyer design
\cite{Swift_ffrsi_07}.
With this system, the projectile was slightly dished on impact,
though this should not affect the pyrometry measurement significantly.
The projectile was accelerating slightly, reverberating elastically
from the acceleration process, and possibly had a porous region through
its thickness as a result of tensile stresses during acceleration.
The relatively strong reverberations in the projectile affect its
effective speed on impact, and contributed to the uncertainty in material
states.

The projectile was Al-6061 alloy, 6\,mm thick, 
accelerated to $3.6\pm 0.1$\,km/s.
The target was Mo, also 6\,mm thick.
Six experiments were performed, four for release into a LiF window 
and two into vacuum.
Surface emission was measured with a 5 channel visible-near infrared pyrometer
or a 4 channel near infrared pyrometer.
The free surface temperature had a relatively large uncertainty,
and the signals on release into LiF showed evidence of thermal emission
from the LiF itself with a temperature of around 580\,K \cite{Seifter_07}.
The measured release temperature was $762\pm 40$\,K into LiF,
and $568\pm 100$\,K from the free surface.

The shocked state in the Mo was calculated to be 
$63.9\pm 2.4$\,GPa and $707\pm 31$\,K,
of which $53\pm 3$\,K was from plastic work.
The state on release into LiF was thus calculated to be
$27.1\pm 1$\,GPa and $670\pm 25$\,K,
of which $89\pm 1$\,K was from plastic work.
The state on release into vacuum was calculated to be
$635\pm 23$\,K,
of which $126\pm 4$\,K was from plastic work.
The uncertainties are correlated: the smallest, mean, and largest of each
go together.

\begin{figure}
\begin{center}\includegraphics[scale=0.7]{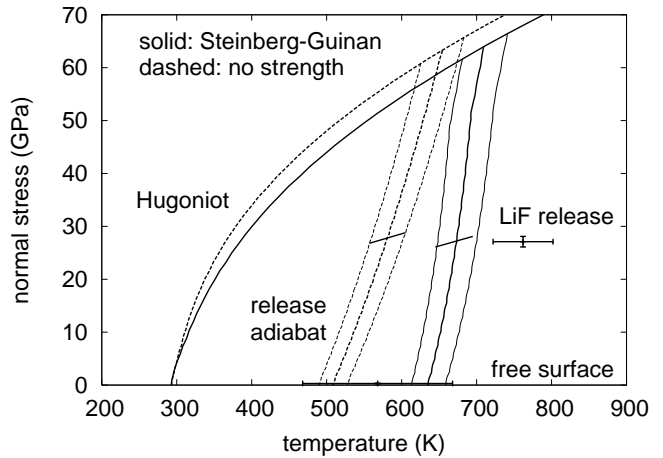}\end{center}
\caption{Temperature measurements from a shock of 64\,GPa,
   on release to 27\,GPa into LiF and to zero pressure,
   compared with predictions based on the Steinberg-Guinan strength model
   and with strength neglected.
   The release adiabat from the mean shock pressure is shown,
   along with adiabats reflecting the uncertainty in shock pressure.
   The lines across the release adiabats show where release pauses when
   a shock is transmitted into the LiF.}
\label{fig:tpcmpff}
\end{figure}

The surface temperature on release into LiF was 1.5-2.5 standard deviations
of the temperature predicted
using the Steinberg-Guinan strength model,
and 3.5-4.5 standard deviations from the temperature predicted
without strength.
The uncertainty in the free surface release temperature was too large to
discriminate between a purely hydrodynamic calculation (no strength)
and the Steinberg-Guinan model -- both lay within one standard deviation
of the measurement.
The predicted temperature of the LiF itself also matched the measurement
to within the experimental uncertainties.
(Table~\ref{tab:states} and Fig.~\ref{fig:tpcmpff}.)

\begin{figure}
\begin{center}\includegraphics[scale=0.7]{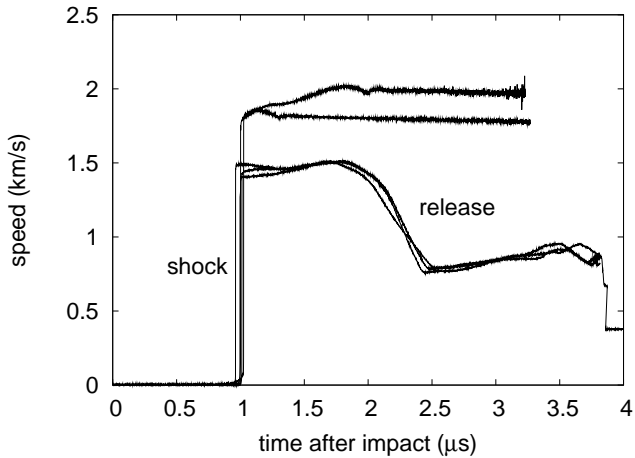}\end{center}
\caption{Surface velocity histories measured in Forest Flyer impact
   experiments with and without a LiF window.
   Separate lines are from different experiments.
   The upper two traces are from free surface release;
   the lower three are from release into a LiF window.}
\label{fig:ffucmp}
\end{figure}

\begin{figure}
\begin{center}\includegraphics[scale=0.7]{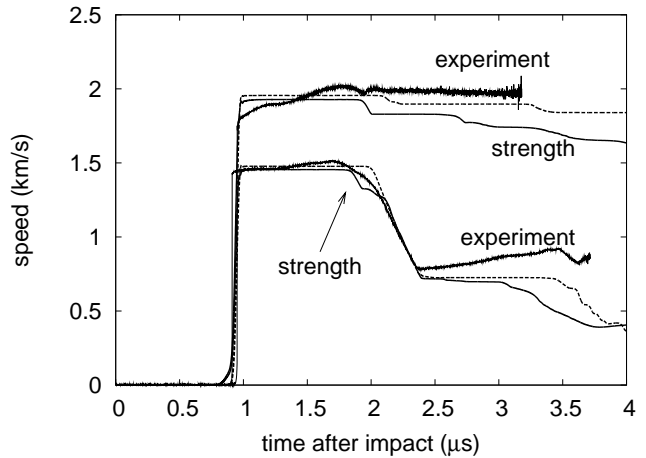}\end{center}
\caption{Surface velocity history in Forest Flyer impact
   experiment with a free surface (upper traces) and a LiF window
   (lower traces).
   Each experimental records is compared with two continuum dynamics simulations,
   with (solid lines) and without strength (dashed lines) in the Mo.
   The elastic precursor to the release wave is evident where the experimental
   velocity histories deviate from the simulations without strength.}
\label{fig:ffucmp1}
\end{figure}

The velocity history of the surface of the sample was measured by
laser Doppler velocimetry of the `VISAR' type \cite{Barker72}.
General features of the velocity history included a rapid 
acceleration when the shock reached the surface,
a roughly constant peak velocity corresponding to the sustained pressure
behind the shock,
deceleration caused by the release wave from the rear of the projectile,
and a slight re-acceleration as the sample was subjected to tensile stress
causing spall type damage (Fig.~\ref{fig:ffucmp}).
The epoch of peak velocity was not perfectly constant, but showed some
acceleration.
This was probably caused by the compression gradient in the projectile from the
residual accelerating pressure at impact, and any regions of porosity resulting
from tensile damage as the projectile was accelerated by the relatively strong
pressures induced by the detonating high explosive.
The onset of release showed a clear elastic precursor
(Fig.~\ref{fig:ffucmp1}).

The measured velocity histories were compared with 
spatially-resolved one-dimensional continuum dynamics simulations.
The projectile was modeled as ideal, i.e. at uniform STP conditions
and traveling at a constant 3.6\,km/s with no reverberations.
As a result, the peak velocity epoch was flatter than measured,
but was in good agreement for amplitude and duration.
Release into the LiF was also reproduced well overall.
The shape of the elastic precursor to release was not reproduced perfectly 
using the Steinberg-Guinan strength model, but its amplitude was reproduced 
to within around 10\%\ and the time of arrival was in good agreement
with the experiment.
The difference in shape could be caused by inadequacy in the Steinberg-Guinan
model -- for example, in the detailed work-hardening history in the
shocked state -- but is more likely to reflect density variations in the
projectile as discussed above.
The uniaxial strains greatly exceeded the elastic limit on release as well
as on compression, so the plastic work should be dominated by the flow
stress rather than the precise path before plastic flow occurred.
Thus the agreement between calculated and observed amplitudes suggests that
the plastic work should be correct to around 10\%.

Spallation did not affect the shock and release states of interest for
the temperature measurements considered here.
The simulations used a crude spall model of the minimum pressure type,
with a minimum pressure of -1.5\,GPa \cite{Steinberg96}, meaning that
the maximum tensile stress induced by the Mo as it was distended
was 1.5\,GPa.
No treatment of accumulating porosity was included, so the Mo as simulated
continued to exert a tensile stress when in reality voids or cracks would
open, reducing the stress.
Thus the simulations of velocity history did not show a re-acceleration 
after the deceleration associated with the release wave.
Tensile damage and spall can depend strongly on the strain rate and
loading history.
The simulated and observed release deceleration matched to within a few percent,
suggesting that the published spall strength applies well to the loading
history induced by these projectile impact experiments.

\begin{table*}
\caption{Shock and release states.}
\label{tab:states}
\begin{center}
\begin{tabular}{|l|c|c|c|c|c|c|c|} \hline
 & \multicolumn{3}{|c|}{no strength} & \multicolumn{3}{|c|}{strength} & measured \\\cline{2-7}
 & particle speed & normal stress & temperature & particle speed & normal stress & temperature & temperature \\
 & (km/s) & (GPa) & (K) & (km/s) & (GPa) & (K) & (K) \\ \hline
gas gun & & & & & & & \\
Mo shock & 0.905 & 57.9 & 594 & 0.902 & 58.7 & 645 & \\
Mo release into LiF & 1.374 & 25.4 & 532 & 1.337 & 24.8 & 614 & $683\pm 41$ \\
LiF shock & 1.374 & 25.4 & 535 & 1.337 & 24.8 & 532 & \\
\hline
Forest Flyer & & & & & & & \\
Mo shock & $0.97\pm 0.03$ & $63.3\pm 2.4$ & $654\pm 28$ & $0.97\pm 0.03$ & $63.9\pm 2.4$ & $707\pm 31$ & \\
Mo release into vacuum & $1.95\pm 0.07$ & 0 & $509\pm 19$ & $1.91\pm 0.07$ & 0 & $635\pm 23$ & $566\pm 100$ \\
Mo release into LiF & $1.48\pm 0.04$ & $27.8\pm 1$ & $581\pm 24$ & $1.44\pm 0.04$ & $27.1\pm 1$ & $670\pm 25$ & $762\pm 40$ \\
LiF shock & $1.48\pm 0.04$ & $27.8\pm 1$ & $570\pm 17$ & $1.44\pm 0.04$ & $27.1\pm 1$ & $566\pm 16$ & $624\pm 100$ \\
\hline\end{tabular}
\end{center}
\end{table*}

\section{Conclusions}
Shock and release temperatures were calculated self-consistently
using the equation of state and a published constitutive model for Mo.
Strength was calculated to make a significant difference to states
in experiments exploring pressures of tens of gigapascals.
The high pressure flow stress predicted using the Steinberg-Guinan
strength model matched the elastic release precursor observed using
surface Doppler velocimetry, suggesting that the flow stress was 
correct to around 10\%.
The predicted temperatures were consistent with pyrometry measurements for
shocks of around 60\,GPa, releasing into a LiF window or into vacuum.
The LiF release temperatures were clearly more consistent with 
plastic work as predicted using the Steinberg-Guinan model than with
hydrodynamic flow (no strength).
The uncertainties in temperature were however too large to discriminate
between strength models to better than several tens of percent in 
flow stress.

Heating from plastic work was calculated to be around 50\,K for shock pressures
around 60\,GPa,
90\,K on subsequent release into LiF, and 125\,K on release at a free surface.
The fraction of plastic work converted to heat was assumed to be 90\%\ -- 
the heating would have been about 10\%\ greater if all the plastic work
appeared as heat.
Taking plastic flow into account, there was no discrepancy between
predictions and measured release temperatures for Mo.
This is a validation of the models of EOS and strength for Mo,
and of the use of pyrometry to measure release temperatures in metals
-- though the pyrometry measurements obtained in these experiments
were not precise enough to discriminate
between models calibrated against similar mechanical data such as velocity
histories.
The fraction of plastic work converted to heat was most likely close to
100\%, though the uncertainty in the temperature measurements means that
this figure cannot be justified statistically to better than a few tens of
percent.

Plastic flow makes a significant contribution to reconciling the temperature
discrepancy observed in the neutron resonance spectrometry experiments on
shocked Mo, although the complete explanation is more complicated and will be
reported separately.

\section*{Acknowledgments}
We would like to acknowledge the contribution of
Carl Greeff for assistance and advice on equations of state and their
uncertainties or certainties for Mo, of
Ron Rabie, David Funk, Rob Hixson, and Chuck Forest 
for detailed information on the design and testing of the 
Forest Flyer loading system, and of Sheng-Nian Luo for general advice and comments
on pyrometry and material dynamics.
The gas gun experiments were performed by D.B.~Holtkamp, P.~Paulsen, P.~Fiske,
D.~DeVore, J.~Garcia, and L.~Tabaka at Lawrence Livermore National Laboratory
in 1999.
The work was performed under the auspices of
the U.S. Department of Energy under contracts W-7405-ENG-36
and DE-AC52-06NA25396.

\end{document}